\documentclass[letter,traditabstract]{aa_my} 
\usepackage{graphicx}
\usepackage{txfonts}
\usepackage{natbib}
\usepackage{rotate}
  \def\ProDiMo{{\sc ProDiMo\ }}
  \def\Ratran{{\sc Ratran\ }}

  \def\Td{T_{\hspace*{-0.2ex}\rm dust}}
  \def\Tg{T_{\hspace*{-0.2ex}\rm gas}}
  \def\cT2{c_T^2}
  
  \def\Rin{R_{\rm in}}
  \def\Rout{R_{\rm out}}

  \def\HH{{\rm\langle H\rangle}}
  \def\nH{n_{\HH}}
  \def\etal{${\rm \hspace*{0.8ex}et\hspace*{0.7ex}al.\hspace*{0.7ex}}$}
  
  \def\ie{i.\,e.\ }
  \def\eg{e.\,g.\ }

  \def\r0{\vec{r}_0}

  \def\amin{{a_{\rm min}}}
  \def\amax{{a_{\rm max}}}
  \def\apow{{a_{\rm pow}}}

\begin{document}

   \title{Hot and cool water in Herbig Ae protoplanetary disks}

   \subtitle{A challenge for Herschel}

   \author{P. Woitke
          \inst{1,2}
          \and
          W.-F. Thi\inst{3}
          \and
          I. Kamp\inst{4}
          \and
          M. R. Hogerheijde\inst{5}
          }

   \institute{UK Astronomy Technology Centre, Royal Observatory, Edinburgh,
              Blackford Hill, Edinburgh EH9 3HJ, UK
         \and
             School of Physics \& Astronomy, University of St.~Andrews,
             North Haugh, St.~Andrews KY16 9SS, UK
         \and
	     SUPA,
             Institute for Astronomy, University of Edinburgh,
             Royal Observatory, Blackford Hill, Edinburgh EH9 3HJ, UK
	 \and
             Kapteyn Astronomical Institute, Postbus 800,
             9700 AV Groningen, The Netherlands
         \and 
	     Leiden Observatory, Leiden University, PO Box 9513, 
             2300 RA Leiden, The Netherlands
             }

   \date{Received April 1, 2009; accepted May 30, 2009}

   \abstract{ The spatial origin and detectability of rotational
     H$_2$O emission lines from Herbig Ae type protoplanetary disks
     beyond 70\,$\mu$m is discussed. We use the recently developed
     disk code \ProDiMo to calculate the thermo-chemical structure of
     a Herbig Ae type disk and apply the non-LTE line radiative
     transfer code \Ratran to predict water line profiles and
     intensity maps. The model shows three spatially distinct regions
     in the disk where water concentrations are high, related to
     different chemical pathways to form the water: (1)~a big water
     reservoir in the deep midplane behind the inner rim, (2)~a belt
     of cold water around the distant icy midplane beyond the
     ``snowline'' $r\!\ga\!20\,$AU, and (3)~a layer of irradiated hot
     water at high altitudes $z/r\!=\!0.1\,...\,0.3$, extending from
     about 1\,AU to 30\,AU, where the kinetic gas temperature ranges
     from 200\,K to 1500\,K. Although region~3 contains only little
     amounts of water vapour ($\sim\!10^{-4}\rm\,M_{Earth}$), it is
     this warm layer that is almost entirely responsible for the
     rotational water emission lines as observable with Herschel. Only
     one ortho and two para H$_2$O lines with the lowest excitation
     energies $<\!100\,$K are found to originate partly from
     region~2. We conclude that observations of rotational water lines
     from Herbig Ae disks probe first and foremost the conditions in
     region~3, where water is predominantly formed via neutral-neutral
     reactions and the gas is thermally decoupled from the dust
     $\Tg\!>\!\Td$. The observation of rotational water lines does
     {\em not} allow for a determination of the snowline, because the
     snowline truncates the radial extension of region~1, whereas the
     lines originate from the region~3.  Different line transfer
     approximations (LTE, escape probability, Monte Carlo) are
     discussed. A non-LTE treatment is required in most
     cases, and the results obtained with the escape probability
     method are found to underestimate the Monte Carlo results by
     2\%$\,-\,$45\%.}


   \maketitle

\section{Introduction}

Water is one of the most important species in planet formation, disk
evolution and for the origin of life. In protoplanetary disks, water
can be abundant either in the gas phase or as solid ice, owing to a
high sublimation temperature.

Water vapour is predicted to be abundant inside of the snowline
within a few AU where densities are high and temperatures are too
warm for ice formation \citep{Agundez2008,Glassgold2009}. Indeed,
observations in the near- and mid-infrared with {\em Spitzer}
\citep{Salyk2008,Carr2008,Najita2009,Eisner2009} reveal the presence
of water vapour and other simple organic molecules (OH, C$_2$H$_2$,
HCN, CO$_2$). Simple modelling of the line emissions constrains the
warm gas (a few 100\,K) to be located inside of $r\!\la\!3\!-\!5$ AU.

The core-accretion model for planet formation is only efficient in the
cold midplane of disks where dust grains are covered by water ice. Icy
grains are stickier than bare silicate grains and coagulate faster
into planetesimals \citep{Ida2008}.  Liquid water is one of the
prerequisites for the emergence of life on terrestrial planets. But
its origin, either via release of water molecules trapped in hydrated
rocks during volcanism or via the impact of comets, is still being
debated \citep{Nuth2008}.

The lowest rotational lines of water lie in the far IR and
are only observable by satellites, \eg the {\em Herschel Space
Observatory}. Other recent works on rotational water lines from
protoplanetary disks used X-ray models (without UV photoprocesses) to
calculate the thermo-chemical disk structure on top of a pre-described
density structure \citep{Meijerink2008b} and applied a multi-zone
escape probability method \citep{Poelman2005,Poelman2006} to compute
the line fluxes.  In this {\em Letter}, we use the disk code \ProDiMo
to compute the disk structure, temperature, and water abundance
self-consistently, and discuss the prospects for detecting rotational
water lines by means of the Monte Carlo code {\sc Ratran}
\citep{Hogerheijde2000}.

\section{The model}

We used the recently developed disk code \ProDiMo to calculate the
thermo-chemical structure of a protoplanetary disk around a Herbig Ae
type star with parameters listed in Table~\ref{tab:Parameter}.
\ProDiMo combines frequency-dependent 2D dust-continuum radiative
transfer, kinetic gas-phase and UV photo-chemistry, ice formation, and
detailed non-LTE heating \& cooling with the consistent calculation of
the hydrostatic disk structure. \ProDiMo does not include X-rays at
the moment. X-ray to FUV luminosity ratios of Herbig Ae stars are
often low $L_{\rm X}/L_{\rm FUV}\!\ll\!1$ \citep{Kamp2008}, so that we
assume that the FUV irradiation provides the main energy input for
the disk. The model is characterised by a high degree of consistency
between the various physical, chemical, and radiative processes, where
the mutual feedbacks are solved by global iterations. For
more details see \citet{Woitke2009}, henceforth called Paper~I.
Recent updates include an improved treatment of UV photorates
by detailed cross sections in the calculated radiation field
\citep[see][]{Kamp2009}.

\begin{figure}
  \centering
  \hspace*{-1mm}\includegraphics[width=8.8cm,height=7.9cm]
                {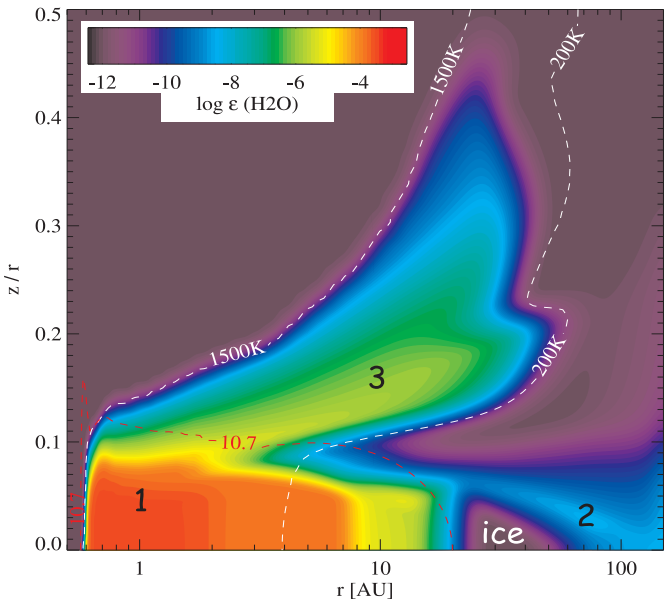}\\[-1mm]
  \caption{Concentration of water molecules $\epsilon_{\rm
    H_2O}\!=\!n_{\rm H_2O}/\nH$ as function of radius $r$ and relative
    height above the midplane $z/r$ for a Herbig Ae disk model. Three
    different regions with high water concentration can be
    distinguished as described in the text.
    The two white contours show $\Tg\!=\!200\,$K and 1500\,K.
    The red contour line marks total hydrogen nuclei
    particle density $\nH\!=\!10^{10.7}\rm\,cm^{-3}$.}
  \label{fig:H2Oconc}
  \vspace*{-1mm}
\end{figure}

\begin{table}
\caption{Herbig Ae type disk model parameter.}
\label{tab:Parameter}
\resizebox{7.6cm}{!}{
\begin{tabular}{l|c|c}
\\[-4.5ex]
\hline
 quantity & symbol & value\\
\hline 
\hline 
stellar mass                      & $M_\star$          & $2.2\rm\,M_\odot$\\
effective temperature             & $T_{\rm eff}$      & $8600\,$K\\
stellar luminosity                & $L_\star$          & $32\rm\,L_\odot$\\
\hline
disk mass                         & $M_{\rm disk}$     & $0.01\rm\,M_\odot$\\
inner disk radius                 & $\Rin$             & 0.5\,AU$^{\,(1)}$\\
outer disk radius                 & $\Rout$            & 150\,AU\\
radial column density power index & $\epsilon$         & 1.0\\
\hline
dust-to-gas mass ratio            & $\rho_d/\rho$      & 0.01\\
minimum dust particle radius      & $\amin$            & $0.1\,\mu$m\\
maximum dust particle radius      & $\amax$            & $200\,\mu$m\\
dust size distribution power index& $\apow$            & 3.5\\
dust material mass density$^{\,(2)}$& $\rho_{\rm gr}$  & 2.5\,g\,cm$^{-3}$\\
\hline 
&&\\[-2.2ex]
strength of incident ISM UV       & $\chi^{\rm ISM}$   & 1\\
abundance of PAHs relative to ISM & $f_{\rm PAH}$      & 0.1\\
\hline
\end{tabular}}\\[1mm]
\hspace*{0mm}\resizebox{6.5cm}{!}{\begin{minipage}{7cm}
\footnotesize
$(1)$: soft inner edge applied, see Woitke\etal (2009)
$(2)$: dust optical constants from \citet{Draine1984}
\end{minipage}}
\vspace*{-4mm}
\end{table}

The model results in a flared disk structure with a puffed-up inner
rim and a vertically extended hot atomic layer above $z/r\!\ga\!0.15\!-\!0.4$
from the inner rim to about $r\!=\!20$\,AU, similar to Fig.~9 in
Paper~I (l.h.s.), where $\Tg\!>\!1000\,$K due to the stellar
UV-irradiation. The major difference between the T\,Tauri type disk
discussed in Paper~I and the Herbig Ae disk discussed here is that in
Herbig disks, the star is much more luminous, so the dust is
warmer in the midplane (here $\Td\!>\!100\,$K inward of
$r\!=\!20\,$AU), which prevents water ice formation. 

Water molecules generally form in deeper layers, and the resulting
water concentration in these layers is depicted in
Fig.~\ref{fig:H2Oconc}. The vertical H$_2$O column densities in this
model are found to be $10^{22}\rm\,cm^{-2}$ at 1\,AU, still
$10^{19}\rm\,cm^{-2}$ at 10\,AU, but then quickly dropping below
$10^{15}\rm\,cm^{-2}$ at 30\,AU and beyond.

\section{Chemical pathways to water}

\begin{table}
\vspace*{-1.2mm}
\caption{Characteristics of water regions in Herbig Ae disk model.}
\vspace*{-2.5mm}
\label{tab:H2Ochara}
\resizebox{\columnwidth}{!}{\begin{tabular}{cccccc}
\hline 
$\!\!$region$\!\!$ & $r$\,[AU] & $z/r$ &
$\!\!\!\!M_{\rm H_2O}\rm\,[M_{Earth}]\!\!\!\!$ & $\langle\Tg\rangle\,$[K]
                               & $\!\!\!\!\langle\Td\rangle\,$[K]$\!\!$\\
\hline
\hline
 &&\\[-2.2ex]
  1   & 0.7$\,-\,$20  & $<\!0.1$      & $0.31$                 &  240
                                                               &  240 \\
  2   &  20$\,-\,$150 & $<\!0.1$      & $3.2\times 10^{\,-5}$  &   38 
                                                               &   37 \\
  3   &   1$\,-\,$30  & 0.1$\,-\,$0.3 & $1.1\times 10^{\,-4}$  &  410
                                                               &  150 \\
\hline
\end{tabular}}
\vspace*{-2mm}
\end{table}

The formation of H$_2$O follows different chemical pathways in the
three different regions shown in Fig.~\ref{fig:H2Oconc}.  Two of these
regions (1 and 3) have been previously identified in vertical slab
models for X-ray irradiated T\,Tauri disks by \citet{Glassgold2009}.

\smallskip\noindent{\sffamily\itshape 1)\ \ The big inner water
reservoir.\ \ } The deep midplane regions from just behind the inner
rim to a distance of about 10\,AU in this model host the majority of
the water in the disk, see Table~\ref{tab:H2Ochara}. This region is
almost completely shielded from the stellar and interstellar radiation
$(A_V\!>\!10)$, is still too warm for water to freeze out $(T_{\rm
dust}\!>\!120\rm\,K)$, and is characterised by large particle densities
$>\!10^{11}\rm\,cm^{-3}$ and thermal equilibrium between gas and dust
($T_{\rm gas}\!\approx\!T_{\rm dust}$). The important chemical
reactions in this region are in detailed balance with their direct
reverse reaction; \ie the gas-phase chemistry is close to
thermochemical equilibrium. Since H$_2$O is the thermodynamically
most stable oxygen carrier under these conditions (after CO), the
water concentration is given approximately by the element abundance of
oxygen minus the fraction thereof bound in CO.

\medskip\noindent{\sffamily\itshape 2)\ \ The distant water belt.\ \ }
Region 2 at $r\!\approx\!20\!-\!150\,$AU and $z/r\!\la\!0.05$ is
characterised by particle densities
$\sim\!10^9\!-\!10^{10}\rm\,cm^{-3}$, temperatures
$\Tg\!\approx\!\Td\!\approx\!30\!-\!100\,$K, and UV-strengths $\chi$
(see Eq.\,(41) in Paper~I) below 500.  These conditions allow water to
freeze out, and there is a tight equilibrium between water adsorption
and photodesorption $\rm H_2O \leftrightarrow H_2O^\#$, where
H$_2$O$^\#$ designates water ice. The formation of
gaseous water in this region is mainly controlled by the following two
photoreactions
\begin{eqnarray}
  \rm H_2O^\# + h\nu &\longrightarrow& \rm H_2O \\ 
  \label{reac1}
  \rm H_2O + h\nu &\longrightarrow& \rm OH + H \ .
  \label{reac2}
\end{eqnarray}
Therefore, the strength of the UV field $\chi$ controls the water
vapour concentration\footnote{Our analysis is restricted to the case
of kinetic chemical equilibrium. Thus, we cannot discuss the history
of water ice formation with our model. Water ice may furthermore be
photodesorbed directly into OH, see e.g. \citet{Andersson2008}.}.  At
$r\!=\!80\,$AU, $\chi$ drops from $\sim\!10^3$ to virtually $0$
between $z/r\!\approx\!0.15$ and $0.02$. The UV photons 
mostly come from the star and are scattered around the optically thick core
of the disk. Water is photo-dissociated for $\chi$ that is too high, and
freezes out for too low $\chi$.  The result is a thin belt of cold
water vapour with concentrations in excess of $10^{-9}$ around the
distant icy midplane.  \citet{Glassgold2009} have not found region 2
in their T\,Tauri models, probably because their model did not include
the UV photoprocesses.

\begin{figure*}
  \centering
  \vspace*{-1mm}
  \begin{tabular}{cccc}
  \hspace*{-2mm}\rotatebox{90} 
  {\hspace*{21mm}\sffamily\itshape\large full model}                & 
    \hspace*{-5mm}\includegraphics[height=5.45cm]{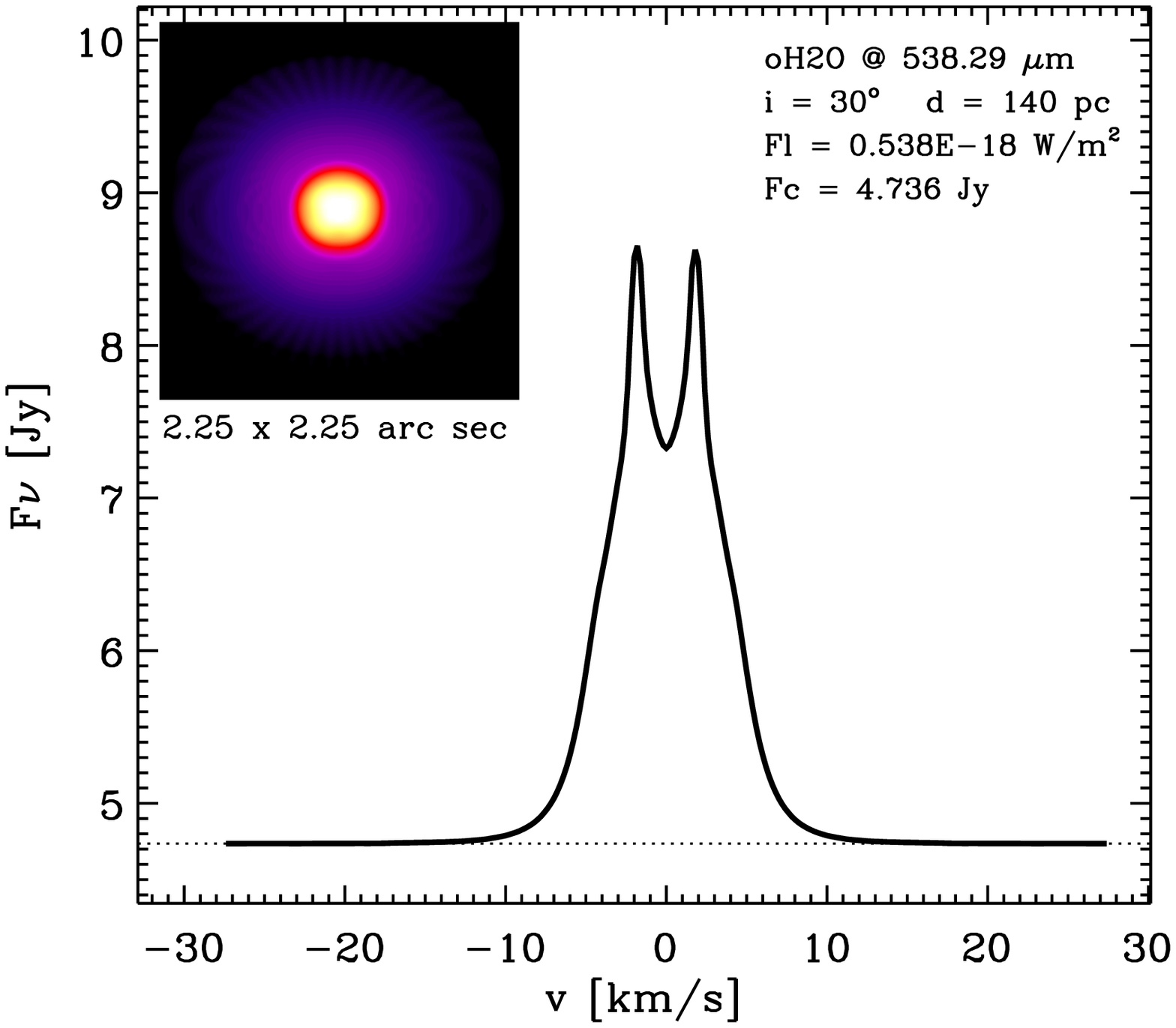} &
    \hspace*{-5mm}\includegraphics[height=5.45cm]{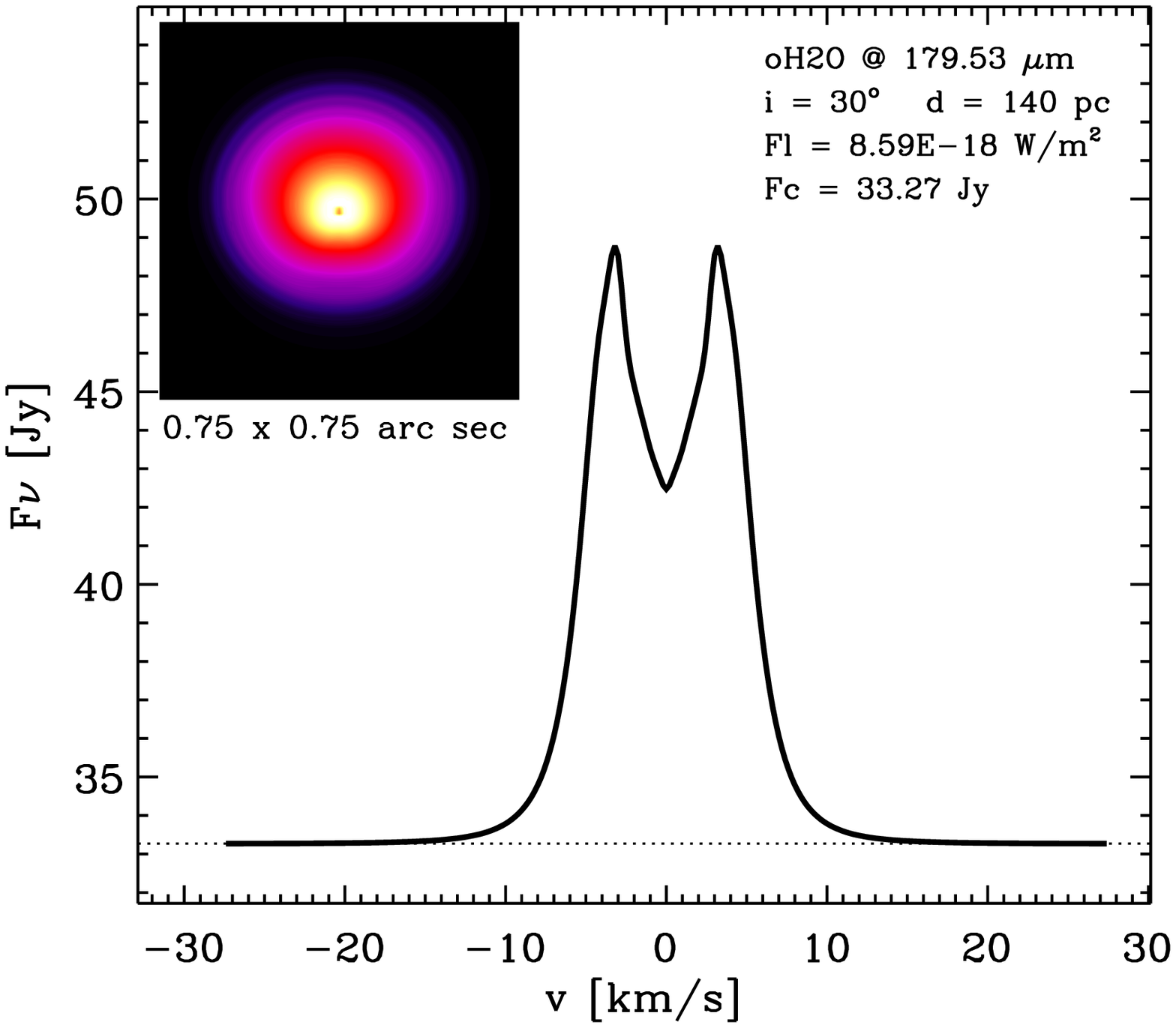} &
    \hspace*{-5mm}\includegraphics[height=5.45cm]{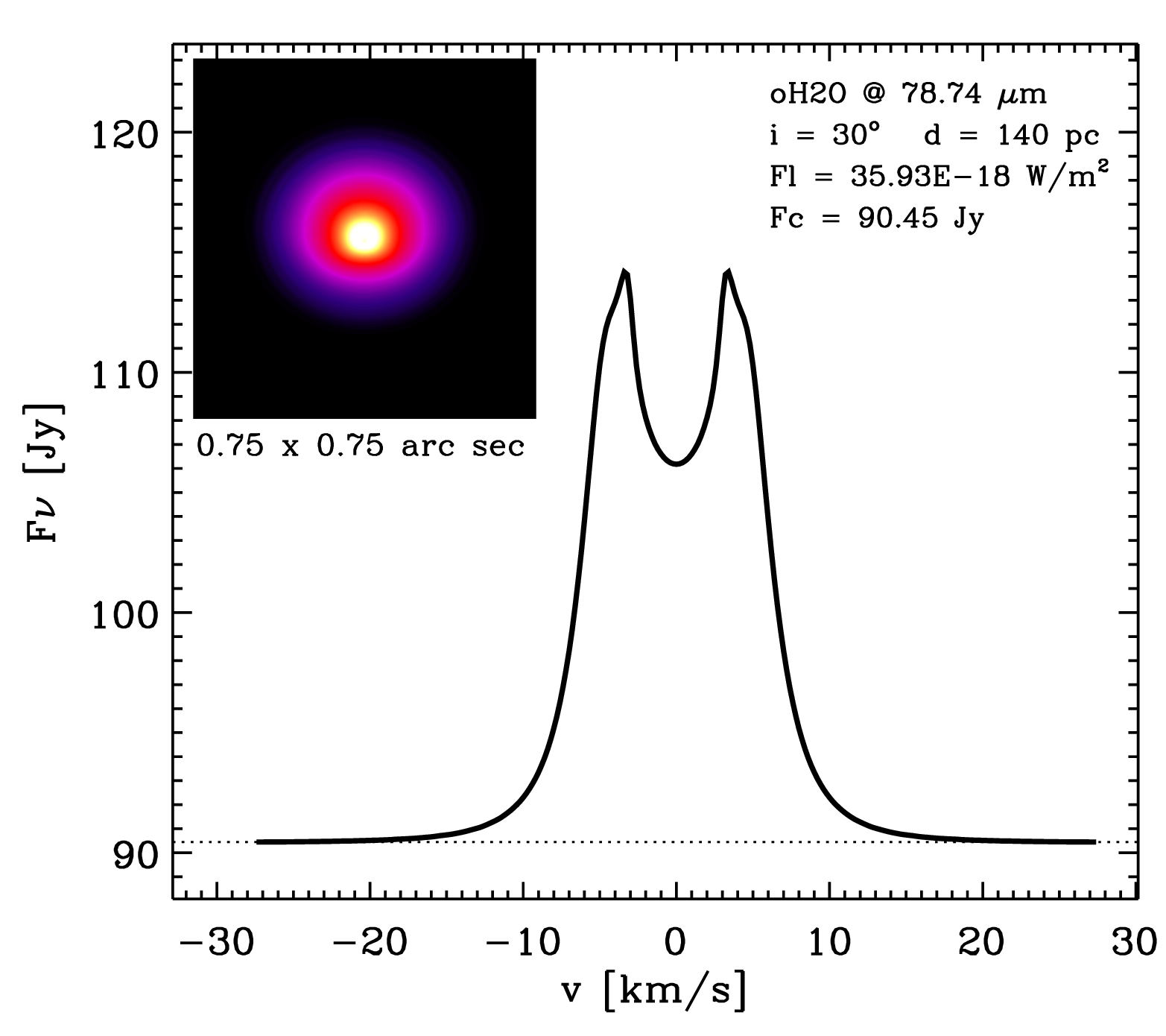}  
    \\[-3mm]
  \hspace*{-1.5mm}\rotatebox{90} 
  {\hspace*{9mm}\sffamily\itshape\large hot water layer 3 removed}  & 
    \hspace*{-5mm}\includegraphics[height=5.45cm]{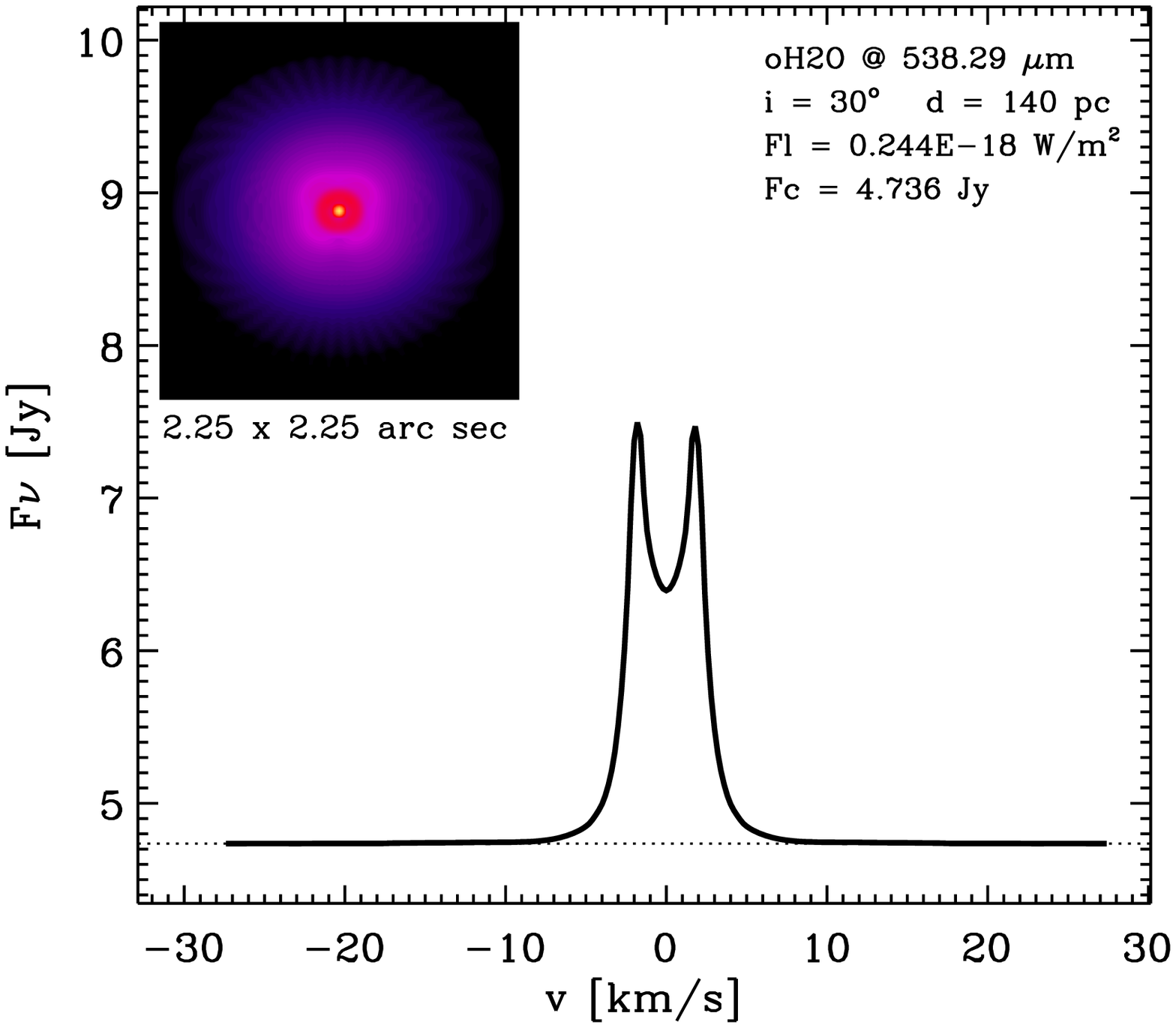} & 
    \hspace*{-5mm}\includegraphics[height=5.45cm]{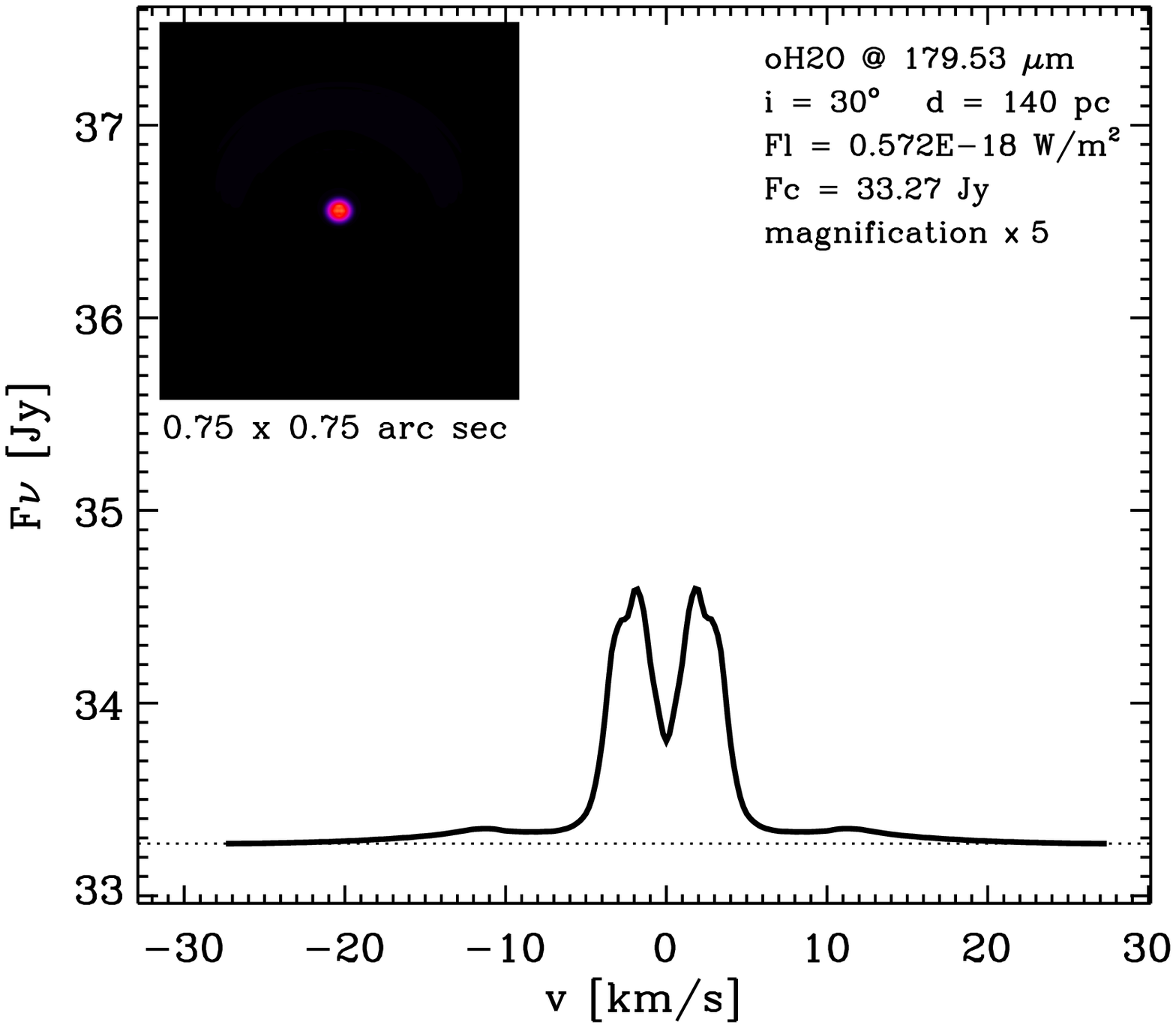} & 
    \hspace*{-5mm}\includegraphics[height=5.45cm]{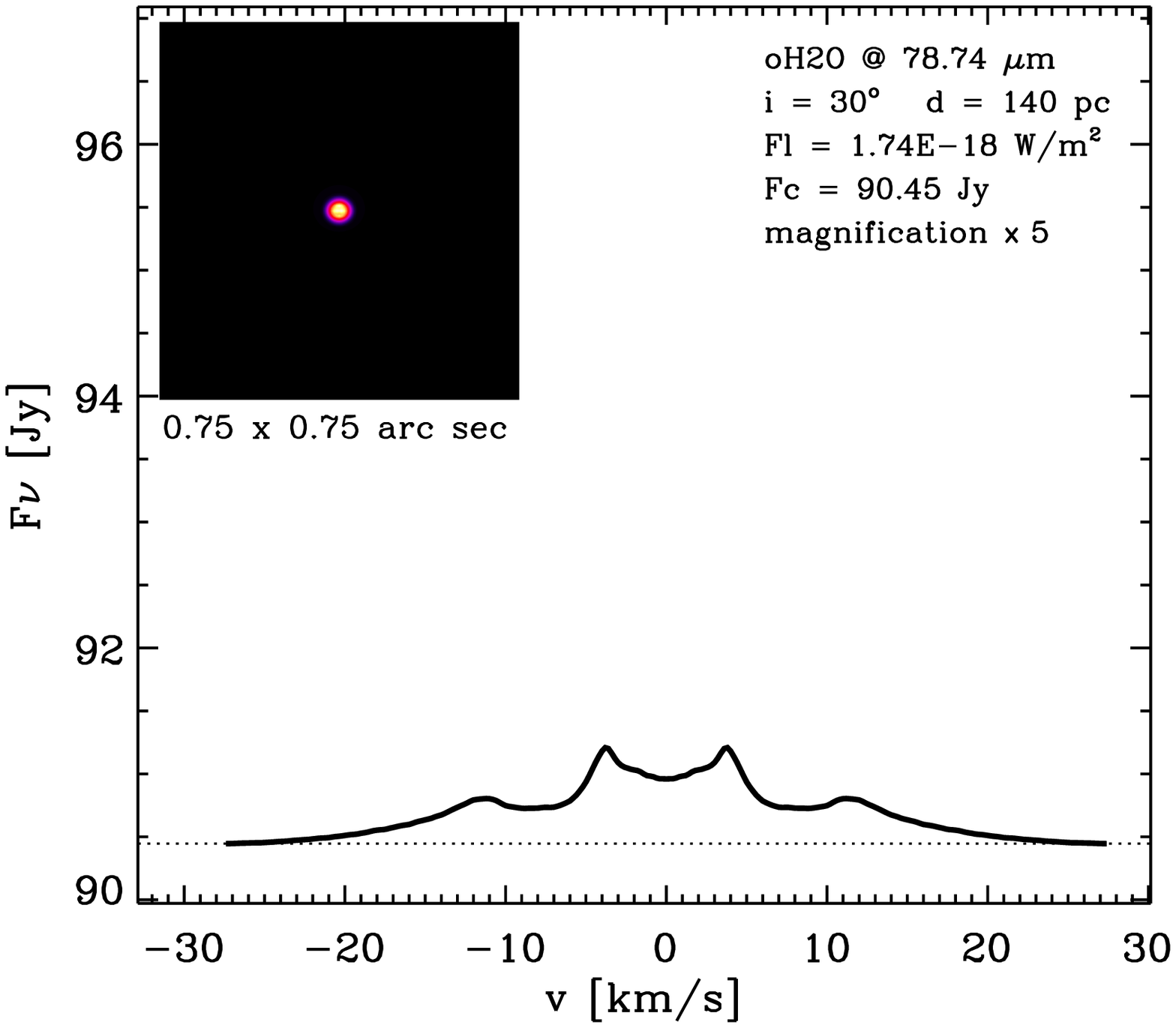}  
  \end{tabular}\\*[-1mm]  
  \caption{Monte Carlo simulations of three ortho H$_2$O lines with
           increasing excitation energy for a distance of 140\,pc
           and inclination $30^\circ$ (infinite resolution).  The
           inserted maps show the integrated line intensity
           $I(x,y)\!=\!\int I_\nu(x,y)-I_\nu^{\rm cont}(x,y)\,d\nu$ as
           function of sky position ($x,y$) with $I_\nu^{\rm
           cont}(x,y)\!=\!I_\nu(x,y)$ far away from line centre.  An
           angular distance of 0.3'' in the maps corresponds to a
           spatial distance of 42\,AU.  The first row shows the
           results for the full thermo-chemical input model. The
           second row shows the results with the water molecules being
           removed from region~3 (see Fig.~\ref{fig:H2Oconc}). The
           low-excitation line on the l.h.s. is depicted on a
           $3\times$ larger angular scale. The scaling of the spectral
           flux $F_\nu$ is partly enlarged by a factor $\times 5$ in
           the lower panels as indicated. The colour-scale for
           the intensity maps is the same in each column.}
  \label{fig:specs}
  \vspace*{-2mm}
\end{figure*}

\medskip\noindent{\sffamily\itshape 3)\ \ The hot water layer.\ \ } At
distances $r\!\approx\!1\!-\!30$\,AU and relative heights
$z/r\!\la\!0.1\!-\!0.3$, the model is featured by an additional layer
of warm water-rich gas that is thermally decoupled from the dust
$\Tg\!>\!\Td$.  The particle density is about
$\sim\!10^8\!-\!10^{10}\rm\,cm^{-3}$ in this layer, the dust
temperature $100\!-\!200$\,K, the gas temperature $200\!-\!1500$\,K,
and the UV radiation field strength $\,\chi\!\approx\!10^4\!-\!10^7$.

Above region~3, the high gas temperatures in combination with the direct
UV irradiation from the star efficiently destroys all OH and
H$_2$O. In region~3, the medium is shielded from the direct stellar
irradiation by the puffed-up inner rim. In the shadow of the inner
rim, $\chi$ drops quickly by about two orders of magnitude (the
remaining UV photons are scattered stellar photons), and water forms
via the following chain of surface and neutral-neutral reactions
\begin{eqnarray}
\rm H + H + dust &\longrightarrow& \rm H_2 + dust\\
\rm H_2 + O      &\longrightarrow& \rm OH + H 
  \label{reac4}\\
\rm OH + H_2     &\longrightarrow& \rm H_2O + H \ ,
  \label{reac5}
\end{eqnarray}
counterbalanced by the photo-dissociation reactions
$\rm OH + hv \rightarrow\rm O + H$ and $\rm H_2O + hv \rightarrow\rm OH + H$.
The reason for the minimum of the H$_2$O concentration between
regions~3 and 1 lies in the high activation energies required for 
reactions (\ref{reac4}) and (\ref{reac5}). These neutral-neutral
reactions become inefficient when the gas
temperature drops below $\Tg\!\la\!200\,$K. The radial
extension of the hot water layer is restricted for the same reason.
Since UV chemistry is lacking in \citep{Glassgold2009}, their water
formation is counterbalanced by charge transfer reactions with H$+$,
which also destroys OH and H$_2$O.


\section{Spectral appearance of rotational water lines}

Having calculated the density structure, the molecular abundances, the
dust and gas temperatures and the continuous radiation field in the
disk, we performed axisymmetric non-LTE line transfer calculations for
selected rotational water lines as summarised in
Table~\ref{tab:linefluxes}. The non-LTE input data for
ortho\,(para) H$_2$O is taken from the Leiden {\sc Lambda} database
\citep{Lambda2005}, which includes 45\,(45) levels, 158\,(157) lines,
and 990\,(990) collisional transitions with H$_2$. A scaled version of
the last data is also applied to collisions with H, which is
essential as region~3 is partly H$_2$-poor and H-rich. The velocity
field is assumed to be Keplerian.
We add a turbulent line width of 0.15\,km/s to the
thermal line width throughout the disk.  The ratio between para and
ortho H$_2$O is assumed to be as in LTE.

We used three different methods of increasing complexity to calculate
the water population numbers: local thermal equilibrium (LTE), a
simple escape probability method (ES, see Sect.~6.1 of Paper\,I) and a
modified version of the 2D Monte Carlo code {\sc Ratran}
\citep{Hogerheijde2000}, see \citep{Kamp2009} for modifications. The
LTE and ES methods used the full 150$\times$150 \ProDiMo
output directly as thermo-chemical input model. The grid size of the MC model
needed to be somewhat reduced for practical reasons. We decided to run
80$\times$80 MC models, which need about $7\times 10^5$ photon
packages to converge to a signal/noise ratio better than 5 for the
worst population number in the worst cell, which takes about 13 CPU
hours on a 2.66\,GHz Linux machine.

To investigate the role of the hot water layer (region~3) for the
spectral appearance of the rotational lines, we calculated two sets
for each model. The first set includes the full chemical input
model. In the second set, we artificially put the water abundance in
region~3 to zero (if $\nH\!<\!10^{10.7}\rm\,cm^{-3}$ and
$\Tg\!>\!200\,$K). The results of the two sets of MC models are
compared in Fig.~\ref{fig:specs}.

We generally observe double-peaked line profiles typical for rotating
gas in emission. In the case of the full input model, the peak separation
generally measures the radial extension of region~3 (Kepler velocity
is 8.8\,km/s at 25\,AU in this model, inclined to $30^\circ$ gives
4.4\,km/s). However, the peak separation of the lowest o-H$_2$O
excitation line (538.29$\,\mu$m) and the two lowest p-H$_2$O
excitation lines (269.27$\,\mu$m and 303.46$\mu$m, see
Table~\ref{tab:linefluxes}) correspond to the full radial extent of
the model, 150\,AU.

The truncated model generally results in $10\!-\!20\times$
smaller line fluxes with a broader, often unclear profile. If the
water in region~3 is missing, the lines originate mainly from region~1,
which is optically thick even in the continuum. Since the gas
is in thermal balance with the dust in region~1 ($\Tg\!\approx\!\Td$),
the lines do not go much into emission in region~1.  Again, the
lowest three excitation lines behave differently, and region~2
contributes by $\sim\!30\%$ for these lines. There is
furthermore one intermittent case (o-H$_2$O 179.5$\,\mu$m with
$E_u\!=\!114$\,K) where the truncated model reveals a
small contribution of the extended region~2 with the character of
low-excitation lines (middle column in Fig.~\ref{fig:specs}).
  
All rotational water lines of the Herbig Ae disk discussed in
this letter are above the $1\sigma$ detection limit of the PACS
spectrometer ($\approx 1\!-\!5\times10^{-18}\rm\,W/m^2$, depending on
$\lambda$). The strongest water lines at $78.74\,\mu$m and
$89.99\,\mu$m are above the $5\sigma$ detection limit. However, the
lines sit on a strong dust continuum, which possibly complicates the
detection by Herschel.

\section{LTE vs. escape probability vs. Monte Carlo}

\begin{figure}
  \centering
  \vspace*{-1mm}
  \hspace*{-2mm}
  \includegraphics[width=7.9cm,height=6.4cm]{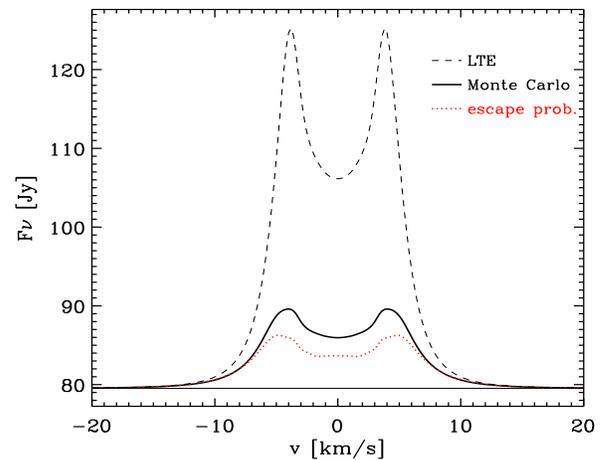}\\[-2mm]
  \caption{Comparison between the results of different line transfer
    methods in application to the high-excitation para-H$_2$O line at
    $89.99\,\mu$m. The full black line redisplays the results of the Monte
    Carlo treatment already shown in Fig.~\ref{fig:specs}, upper row,
    middle figure.}
  \label{fig:LineMethods}
  \vspace*{-1mm}
\end{figure}

Figure~\ref{fig:LineMethods} compares the results obtained by three
different line transfer methods for the high-excitation para-H$_2$O
line at 89.99$\,\mu$m. The results for the other lines are listed in
Table~\ref{tab:linefluxes}. We consider the deviations from the
results of the most advanced method (MC) as a measure of the quality
of the other methods. The LTE predictions are generally too high, by
up to a factor of 3.5, although continuum flux, line width, and peak
separation are similar. Since the densities in region~3 are lower
than the critical density $\sim\!10^{10}\rm\,cm^{-3}$, the levels tend
to depopulate radiatively, which explains the overpredictions by LTE.
Deviations between ES and MC are between 2\% and 45\%, and increase
with excitation energy $E_u$. Our ES method tends to underestimate the
line fluxes in general. The levels in region~3 are pumped by line
radiation from distant regions in LTE that have larger line source
functions. This effect is difficult, if not impossible, to be properly
account for in the ES approximation.


\begin{table}
\caption{Properties of rotational water lines, and calculated
  line fluxes $F_L\!=\!\int (F_\nu-F_\nu^{\rm cont})\,d\nu$ for
  different line transfer methods. Listed line fluxes are
  in units $10^{-18}\rm\,W/m^2$ for distance 140\,pc and
  inclination $30^\circ$. $^\star$ indicates that this
  line will be observed by the Herschel OT Key Programme {\sc
  Gasps}. ES$\,=\,$escape probability, MC$\,=\,$Monte Carlo.}
\label{tab:linefluxes}
\resizebox{\columnwidth}{!}{\begin{tabular}{clp{8mm}cc|cccr}
\\[-4ex]
\hline
\\[-2.3ex]
$\!\!$carrier  & transition$\!\!$ & $\lambda\,[\mu$m] 
               & $\!\!\!A_{ul}\rm\,[s^{-1}]\!\!\!$ 
         & $\!\!\!E_u\rm\,[K]\!\!\!$ & LTE & ES & MC 
                                 & $\!\!\!\!|$ES-MC$|\!\!\!\!$ \\
\\[-2.3ex]
\hline
\hline
\\[-2.3ex]
$\!\!$o-H$_2$O & $1_{10}$$\,\to\,$$1_{01}$ & 538.29 & 0.0035 & 61  
    & $\!\!$0.702$\!\!$ & 0.531 & $\!\!$0.538$\!\!\!\!$  &  2\%\\
$\!\!$o-H$_2$O & $4_{23}$$\,\to\,$$3_{12}$$^\star\!\!\!\!$ 
                                           & 180.49 & 0.0306 & 194 
                           & 9.27  & 2.32   & 3.39$\!\!$ & 37\%\\
$\!\!$o-H$_2$O & $2_{12}$$\,\to\,$$1_{01}$$^\star\!\!\!\!$ 
                                           & 179.53 & 0.0559 & 114 
                           & 11.9  & 7.69   & 8.59$\!\!$ & 11\%\\
$\!\!$o-H$_2$O & $3_{03}$$\,\to\,$$2_{12}$ & 174.63 & 0.0505 & 197 
                           & 12.7  & 7.47   & 8.88$\!\!$ & 17\%\\
$\!\!$o-H$_2$O & $2_{12}$$\,\to\,$$1_{10}$ & 108.07 & 0.256  & 194 
                           & 54.8  & 11.4   & 17.9$\!\!$& 44\%\\
$\!\!$o-H$_2$O & $4_{23}$$\,\to\,$$3_{12}$$^\star\!\!\!\!$ 
                                           & \ 78.74 & 0.484  & 432 
                           & 94.5   & 24.6  & 35.9$\!\!$ & 37\%\\
\hline
\\[-2.2ex]
$\!\!$p-H$_2$O & $2_{02}$$\,\to\,$$1_{11}$ & 303.46 & 0.0058 & 101 
                           & 1.27  & 1.48   & 1.40       &  5\%\\
$\!\!$p-H$_2$O & $1_{11}$$\,\to\,$$0_{00}$ & 269.27 & 0.0184 & 53  
                           & 2.40  & 2.22   & 2.09       &  6\%\\
$\!\!$p-H$_2$O & $4_{13}$$\,\to\,$$3_{22}$$^\star\!\!\!\!$
                                           & 144.52 & 0.0332 & 397 
                           & 5.53  & 4.57   & 4.15$\!\!$ &  9\%\\
$\!\!$p-H$_2$O & $3_{13}$$\,\to\,$$2_{02}$ & 138.53 & 0.125  & 204 
                           & 14.3  & 7.76   & 8.71$\!\!$ & 12\%\\
$\!\!$p-H$_2$O & $2_{20}$$\,\to\,$$1_{11}$ & 100.98 & 0.260  & 196 
                           & 32.5  & 8.66   & 10.9$\!\!$ & 23\%\\
$\!\!$p-H$_2$O & $3_{22}$$\,\to\,$$2_{11}$$^\star\!\!\!\!$
                                           & \ 89.99 & 0.352  & 297 
                           & 48.0  & 10.2   & 14.0$\!\!$ & 31\%\\
\hline
\end{tabular}}
\vspace*{-1.5mm}
\end{table}

\section{Conclusions}

The rotational water lines from Herbig Ae disks beyond 70$\mu$m
originate predominantly from a warm molecular layer at relative
altitudes $z/r\!\approx\!0.1\!-\!0.3$ where H$_2$O is formed via
neutral-neutral reactions in a thermally decoupled gas
($\Tg\!>\!\Td$). The more distant cold water around the icy
midplane, where the ice is photodesorbed, contributes only to the
lowest excitation lines. The peak separation of all other lines
measures the radial extension of the warm molecular layer, which is
about 40\,AU in the discussed model.  In contrast, the vast majority
of water vapour in the disk is situated in the deep midplane,
extending from just behind the inner rim outward to the snowline,
where water freezes out to form water ice. The gas in this massive
deep water reservoir is in thermal balance ($\Tg\!\approx\!\Td$) with
optically thick dust and, therefore, no strong line emissions are
produced with respect to the continuum from this deep region. Thus, no
information about the position of the snowline can be deduced from
the rotational water lines. A similar conclusion was reached by
\citet{Meijerink2008b}. The line analysis generally requires a non-LTE
treatment. Our escape probability method is found to underestimate 
the water line fluxes with respect to the more expensive Monte Carlo method 
by about 2\%$\,-\,$45\%.

\begin{acknowledgements}
We thank Dr.~Rowin Meijerink for an open discussion about water in 
disks and Dr.~Dieter Poelman for internal benchmark tests of
different non-LTE line transfer methods.
\end{acknowledgements}

\bibliography{reference}


\end{document}